\documentclass[a4paper,twoside]{article}

\usepackage{epsfig}
\usepackage{subcaption}
\usepackage{calc}
\usepackage{amssymb}
\usepackage{amstext}
\usepackage{amsmath}
\usepackage{amsthm}
\usepackage{multicol}
\usepackage{pslatex}
\usepackage{apalike}

\usepackage{graphicx} 
\usepackage{url} 
\usepackage{mathtools, cuted} 
\usepackage{booktabs} 
\usepackage[justification=centering]{caption} 

\usepackage{SCITEPRESS}     

\begin{document}

\title{Upgraded W-Net with Attention Gates and its Application in Unsupervised 3D Liver Segmentation}

\author{\authorname{Dhanunjaya Mitta\sup{1}, Soumick Chatterjee\sup{1,2,3}\orcidAuthor{0000-0001-7594-1188}, Oliver Speck\sup{3,4,5,6}\orcidAuthor{0000-0002-6019-5597} and Andreas Nürnberger\sup{1,2}\orcidAuthor{0000-0003-4311-0624}}

\affiliation{\sup{1}Faculty of Computer Science, Otto von Guericke University, Magdeburg, Germany}
\affiliation{\sup{2}Data and Knowledge Engineering Group, Otto von Guericke University, Magdeburg, Germany}
\affiliation{\sup{3}Biomedical Magnetic Resonance, Otto von Guericke University Magdeburg, Germany}
\affiliation{\sup{4}German Center for Neurodegenerative Disease, Magdeburg, Germany}
\affiliation{\sup{5}Center for Behavioral Brain Sciences, Magdeburg, Germany}
\affiliation{\sup{6}Leibniz Institute for Neurobiology, Magdeburg, Germany}
\email{mitta.dhanunjaya@gmail.com, \{soumick.chatterjee, oliver.speck, andreas.nuernberger\}@ovgu.de}
}


\abstract{Segmentation of biomedical images can assist radiologists to make a better diagnosis and take decisions faster by helping in the detection of abnormalities, such as tumors. Manual or semi-automated segmentation, however, can be a time-consuming task. Most deep learning based automated segmentation methods are supervised and rely on manually segmented ground-truth. A possible solution for the problem would be an unsupervised deep learning based approach for automated segmentation, which this research work tries to address. We use a W-Net architecture and modified it, such that it can be applied to 3D volumes. In addition, to suppress noise in the segmentation we added attention gates to the skip connections. The loss for the segmentation output was calculated using soft N-Cuts and for the reconstruction output using SSIM. Conditional Random Fields were used as a post-processing step to fine-tune the results. The proposed method has shown promising results, with a dice coefficient of 0.88 for the liver segmentation compared against manual segmentation.}

\onecolumn \maketitle \normalsize \setcounter{footnote}{0} \vfill

\section{\uppercase{Introduction}}
\label{sec:introduction}

\noindent Image Segmentation is the process of dividing an image into multiple segments, where the pixels in each segment are connected with respect to their intensities or by Regions of Interest \cite{anjna2017review}. Segmentation of biomedical images is a major advancement in the field of medical imaging, as it helps radiologists and doctors to make better and faster decisions. Many approaches to medical image segmentation using various deep learning techniques have been proposed. These methods, however, require a large amount of training data with their respective segmentation masks also known as ground truth images \cite{badrinarayanan2017segnet,chaurasia2017linknet,krahenbuhl2011efficient,paszke2016enet,zheng2015conditional}. Abdominal MR image segmentation is an interesting and challenging research area \cite{gotra2017liver}, but not yet very much explored, until recently \cite{data}. While performing abdominal segmentation, liver segmentation is one of the most challenging task due to the high variability of its shape and its proximity to various other organs \cite{gotra2017liver}. This research addresses the challenge of segmenting the liver from 3D MR images without using any manual ground truth for training the deep neural network model. \\
\indent Our state of the art model is based on W-Net \cite{xia2017w} with both the U-Nets replaced by Attention U-Nets \cite{oktay2018attention}. The original W-Net works with 2D images, but as we want to work with volumetric 3D MR images, the network  architecture was adapted for 3D images by using 3D convolution layers (Sect. 
\ref{sec:model}) and by modifying the calculation of pixel weights to voxel weights (Sect. \ref{sec:preprocessing}). We show the applicability of our approach for liver segmentation using the CHAOS \cite{data} challenge dataset (Sect. 
\ref{sec:dataset}).
 
\subsection{Related work}
\noindent Many approaches to image segmentation have been proposed by different researchers. A variety of atlas-based segmentation methods have been described \cite{gee1993elastically,crum2001automated,baillard2001segmentation}. \textit{Aganj et.al} introduced an approach by computing the local center of mass of the putative region of each pixel, to perform unsupervised medical image segmentation \cite{aganj2018unsupervised}. \textit{Dong Nie et.al} proposed an approach for brain image segmentation of infants by using deep neural networks \cite{nie2016fully}. \textit{Christ et. al } designed an approach by joining two fully cascaded neural networks for automatic segmentation of the liver and its lesions in low-contrast heterogeneous medical volumes \cite{christ2017automatic}. \textit{Oktay et.al} introduced a novel attention gate, which will implicitly learn to suppress regions that are not relevant \cite{oktay2018attention}. These gates are applied to the standard U-Net architecture to highlight the important features, which are passed through skip connections. Noise and irrelevant information in skip connections are eliminated by extracting coarse-scale information in gating. This is performed right before the concatenation operation to merge only relevant activations. \textit{Xide Xia et.al} proposed an approach for a W-Net model by stacking two U-Nets one after another, for unsupervised image segmentation, but for non-medical RGB images \cite{xia2017w}. By using this model, segmentation maps can also be predicted even for applications, which do not have any labeling information available.

\subsection{Contribution}
\noindent Most of the research on biomedical image segmentation using deep learning by now has been focused on supervised learning. This research is a proof-of-concept for biomedical image segmentation using unsupervised learning. The current results are not perfect, but there are many scopes for improvements - that will be discussed later. In this research, a novel 3D Attention W-Net architecture has been proposed, which has been built by replacing the 2D U-Nets of the original W-Net \cite{xia2017w}, by the 3D Attention U-Nets \cite{oktay2018attention}, and for the reconstruction loss, SSIM \cite{larkin2015structural} has been used. Furthermore, some minor changes were introduced to the Attention U-Net architecture before incorporating them to the W-Net, which are discussed in a later chapter. 


\section{\uppercase{Methodology}}
\subsection{Dataset}
\label{sec:dataset}
\noindent The dataset that has been used in this study has been provided by the CHAOS Challange \cite{CHAOSdata2019,data}. The dataset consists of a CT Dataset of 40 subjects, and an MRI Dataset of 40 subjects, with two different sequences - T1-DUAL and T2-SPIR. T1-DUAL contains in-phase and opposed-phase images. For our work, we choose the available 40 volumes of T1-DUAL in-phase. The dataset came with a manually labeled ground-truth. For the purpose of this research they were intentionally ignored during training. Those ground-truths were used only during the evaluation of the algorithm's performance.  

\subsection{Pre-processing}
\label{sec:preprocessing}
\noindent The images were normalized to have pixel values between (0,1) before supplying them to the network, to bring them to a common scale for faster convergence while training. Simultaneously, the weights between the voxels were calculated using Eq. \ref{eq:weight}, where $w_{ij}$ is the weight between the pixel i and j, which is required in calculating Normalized-Cuts using Eq. \ref{eq:softncuts} (loss function). The architecture is based on auto-encoders in which the encoder part maps the input to the pixel-wise segmentation layer without losing its original spatial size and the decoder part reconstructs the original input image from the dense prediction layer.

\begin{multline}\label{eq:weight}
    \omega _{i,j} = e^{\frac{-\left \| F_{i}-F_{j}\right \|^{2}}{\sigma _{I}^{2}}}*\left \{ \begin{array}{rcl}
e^{\frac{-\left \| X_{i}-X_{j} \right \|^{2}}{\sigma _{X}^{2}}} &\end{array} \right. 
\end{multline}

\subsection{Model Construction: 3D Attention W-Net}\label{sec:model}
The base W-Net architecture proposed by \cite{xia2017w} was modified by replacing both the U-Nets with 3D Attention U-Nets \cite{oktay2018attention}. The original W-Net was proposed for 2D Images, both weight calculation and soft ncuts loss calculation have been adopted for 3D.

\begin{figure*}
     \begin{center}
    	\includegraphics[width=\textwidth]{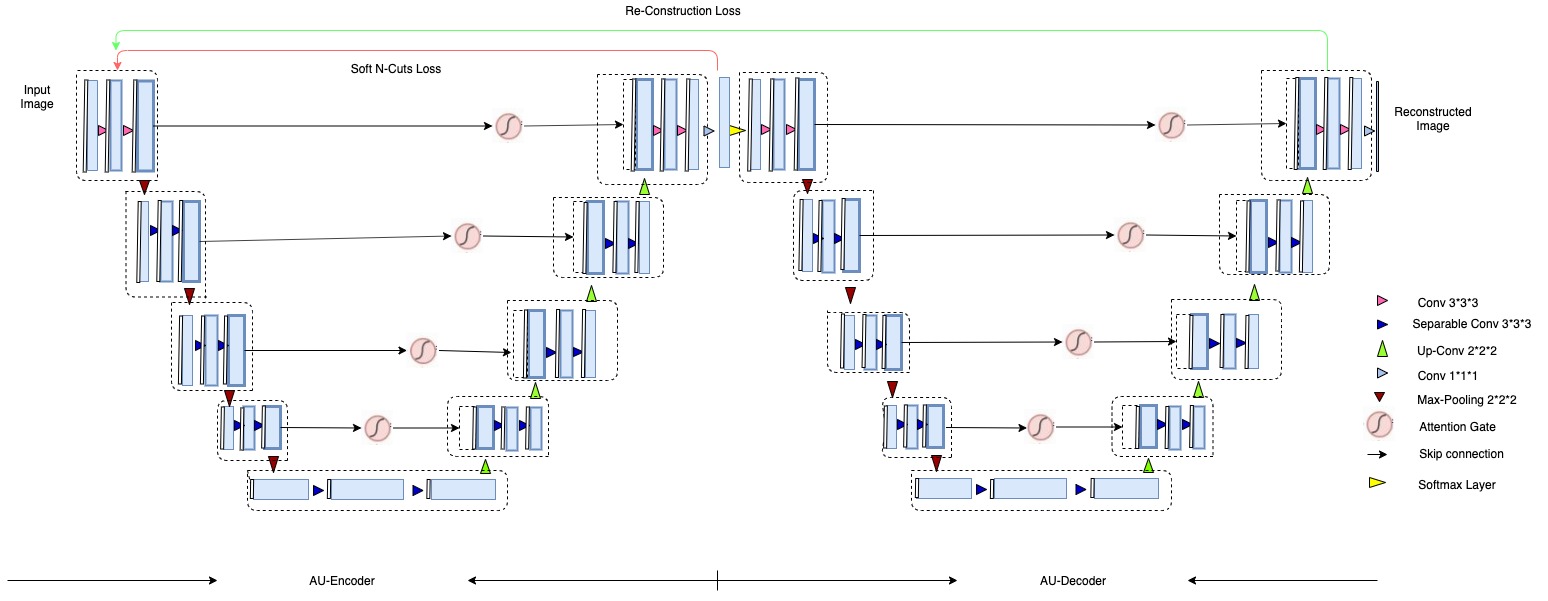}
    	\caption{3D Attention W-Net}
    	\label{fig:W-Net}
     \end{center}
    \end{figure*}

The network is illustrated in \ref{fig:W-Net}. The network consists of two parts.

\begin{itemize}
    \item AU-Encoder, which is on the left side of the network and 
    \item AU-Decoder, which is on the right.
\end{itemize}

The network consists of 18 modules (marked with dotted lines in Figure \ref{fig:W-Net}), each module consists of two 3D convolutional layers with kernel size three. Each layer is followed by a non-linear activation function and an instance normalization layer. In total, we are using 46 3D convolutional layers. The first nine modules represent the encoder network which predicts the segmentation maps and the next nine modules reconstruct the original input image from the segmentation output coming from the encoder part. 

The most frequently used non-linear activation function is the Rectified Linear Unit (ReLU). However, there is a chance of dying neurons \cite{lu2019dying}. Therefore, we used the Parametric Rectified Linear Unit or PReLU \cite{he2015delving}, which is similar to LeakyReLU with the difference of using the hyper-parameter $\alpha$ for negative results, which is adaptively learnt during the training, instead of using a fixed value (such as 0.01) as in LeakyReLU. The data used in this research contain different patient data and the number of slices differs between subjects. To construct batches, data padding to an equal number of pixels would be required. Instead, we used a batch size of one while training the network.

As mentioned in the literature \cite{oktay2018attention}, the encoder part consists of a contracting path that captures context and an expansion path that enables precise localization. As shown in Figure \ref{fig:W-Net}, an input image is given to the first module of the encoder part. Then it undergoes convolutional operations followed by PReLU and instance normalization twice before moving forward to the next module. The modules are connected through 3D max pooling layers, which decrease the image size by two. We also store the original image size before performing the pooling operation recover the image size during the expansion path of the U-Nets. The initial module produces 64 feature maps as output and after every module, the number of features is increased by two. 

\begin{figure}
     \begin{center}
    	\includegraphics[width=0.48\textwidth]{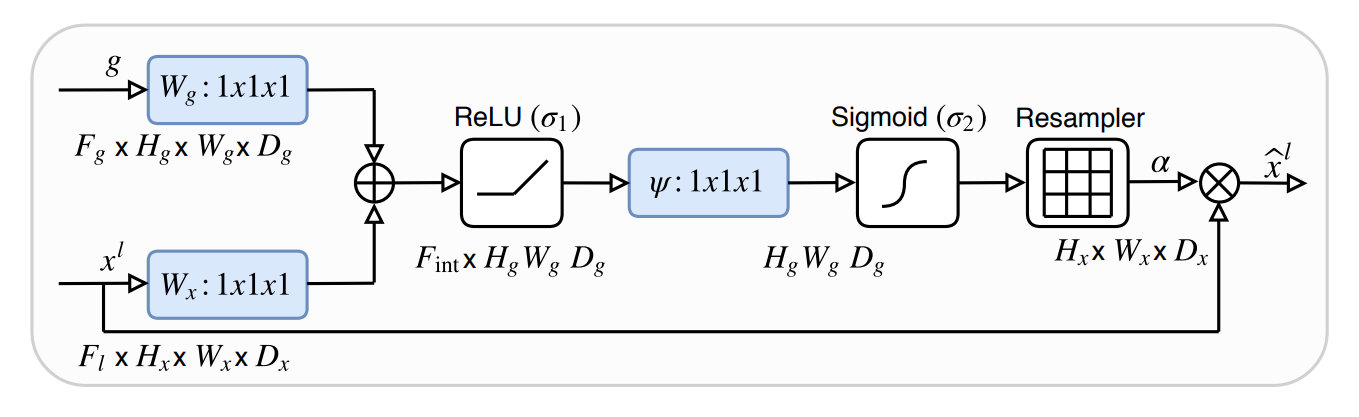}
    	\caption{Attention gate \cite{oktay2018attention} 
    	}
    	\label{fig:attention_gate}
     \end{center}
    \end{figure}

In the contraction path, modules are connected via max pool, which is indicated in brown color; in the expansion path, modules are connected through the upsample layer followed by modules similar to the contraction path and are denoted with green color arrows. Upsampling is performed using trilinear interpolation, and the output size of the interpolation is set to the image sizes saved ptior to each of the max pool operations. Skip connections are passed through attention gates to suppress irrelevant regions and noisy responses. The attention gate architecture is from \cite{oktay2018attention} and shown in Figure \ref{fig:attention_gate}.

The output of the encoder is passed to a fully connected 3D convolution layer with a kernel size of one and a stride of one, followed by a softmax layer. This convolution layer helps to map the 64 feature maps of the output to the required number of K classes, and the softmax function rescales them to (0,1) with a summation of all the K feature maps as one. During the inference stage, the output of the softmax layer is the final output of the model. During training, the output of the softmax is given as the input of the first module of the second U-Net.
The second U-Net is similar to the first one, with the only differences being the final fully connected convolution layer and the final activation function. The fully connected layer provides one final output instead of K outputs. For the final activation function, the sigmoid function is used instead of the softmax, which also rescales the output between (0,1) but doesn't make the sum equal to one. 

\subsection{Loss Functions}
We used two loss functions in this research. The first one is directly after the encoder U-Net, to optimize the encoder U-Net only; the other one is at the end of the decoder U-Net to optimize both the U-Nets.
\subsubsection{N-Cuts Loss}
The first loss function applied to the output of the encoder U-Net is the N-Cuts loss \cite{xia2017w}. The output from the softmax layer of the encoder U-Net is a K-class prediction for each voxel. Normalized cuts from \cite{shi2000normalized} as a global criterion for image segmentation, as shown in Eq. \ref{eq:ncuts} are applied, where $A_{k}$ is the number of voxels in segment k, V is the total number of voxels, and w calculates the weight between two pixels.

Since the argmax function is non-differentiable, it is not possible to get the corresponding gradients during back-propagation. Therefore, the soft n-cuts loss \cite{ghosh2019understanding} is used as shown in Eq. \ref{eq:softncuts}, where $p(u = A_{k})$ measures the probability of node u belonging to class $A_{k}$. The output of the encoder U-Net is forwarded to this soft N-Cuts loss function along with the voxel-weight calculated during the pre-processing stage, following Eq. \ref{eq:weight}. The network is trained to minimize the N-Cuts loss, by optimizing the parameters of the encoder U-Net.

\begin{equation}\label{eq:ncuts}
    Ncut_{k}(V) = \sum_{k = 1}^{K}\frac{\sum _{u\epsilon A_{k}, v\epsilon (V-A_{k})}\omega (u,v)}{\sum _{u\epsilon A_{k}, t\epsilon (V)}\omega (u,t)}  
\end{equation}
\begin{multline}\label{eq:softncuts}
    J_{soft-Ncut}(V,K) = \\  K - \sum_{k = 1}^{K}\frac{\sum _{u\epsilon V}p(u = A_{k})\sum _{u\epsilon V}\omega (u,v)P(v = A_{k})}{\sum _{u\epsilon V}p(u = A_{k})\sum _{t\epsilon V}\omega (u,t)}  
\end{multline}

\subsubsection{Reconstruction Loss}
Reconstruction loss is used to calculate the loss between the output of the decoder U-Net and the input image. The network was trained to minimize the reconstruction loss similar the auto-encoder architecture. Structural Similarity Index (SSIM) is used to calculate the reconstruction loss. A higher SSIM, however, is better and thus the negative of the SSIM value has been used.The network was trained to minimize the reconstruction loss, by optimizing the parameters of both the U-Nets.

SSIM is used to measure the similarities within the pixels i.e., whether the pixels in the images those are being compared have similar pixel density values. SSIM values lie between (0,1), where 1 indicates that both images are identical.

We calculate SSIM by using the following formula:

\begin{multline}\label{eq:ssim}
    \textit{SSIM}(x,y) = \frac{(2\mu_x\mu_y + C_1) + (2 \sigma _{xy} + C_2)} 
    {(\mu_x^2 + \mu_y^2+C_1) (\sigma_x^2 + \sigma_y^2+C_2)} 
\end{multline}
    
where $\mu_x$, $\mu_y$ are the mean values of x and y, $\sigma_x^2$, $\sigma_y^2$ are the variance of x and y, $\sigma _{xy}$ is the co-variance of x and y, $C_1$, $C_2$ are two variables to stabilize the division with weak denominator.

The network was trained with the MR abdominal dataset provided by \cite{data}, which contains 40 volumes. The given set was divided into a training (25 volumes), a validation (5 volumes), and a test set (10 volumes). Both N-cuts and reconstruction loss were minimized, given equal priority (weights) to both loss functions. 

\subsection{Post-processing using Conditional Random Fields}
The use of many max-pooling layers may result in increased invariance, which can cause localization accuracy reduction. To obtain fine boundaries in the output segments, conditional random fields or CRF \cite{chen2017deeplab} were applied as a post-processing step in a 3D CRF variant \cite{chatterjee2020}.

\begin{equation}\label{eq:crf}
    E(X)=\sum \phi (u) + \sum \psi (u,v)
\end{equation}

Where u and v are the voxels, $\phi (u)$ is the unary potential and $\psi (u,v)$ is the pair-wise potential.

After the CRF,  the cluster values corresponding to the liver were identified manually for one volume. The selected clusters were merged to obtain the liver segmentation for the remaining volumes. 
\section{\uppercase{Results}}
We used two U-Nets to form a W-Net for training the model on a given training dataset; during testing, only the first U-Net was used, as the output of the first U-Net corresponds to the automatic segmentation. This predicted segmentation was then passed through the CRF post-processing to recover the boundaries. The network was trained to predict 15 different clusters to segment various parts of the image. Then, the clusters with the liver segment were identified as the final result. The relevant cluster numbers were chosen from only one test volume and applied to all other test volumes. 

\begin{figure*}[ht!]
\centering
\includegraphics[width=0.3\textwidth]{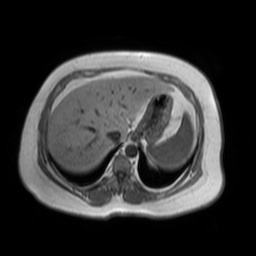}
\includegraphics[width=0.3\textwidth]{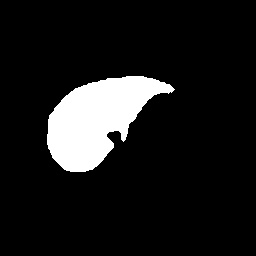}
\includegraphics[width=0.3\textwidth]{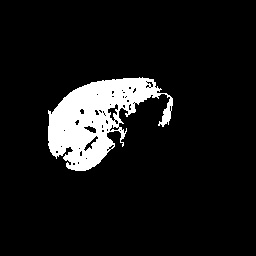}
\caption{Example slice of a test volume: From left to right - Original slice, Ground Truth, only the liver segment, Output of the network - Clusters containing the liver segmentation were considered as our region of interest. \centering}        
\label{fig:seg_1}
\end{figure*}

\begin{figure*}[ht!]
\centering
\includegraphics[width=0.3\textwidth]{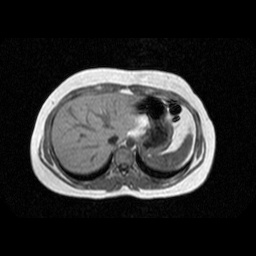}
\includegraphics[width=0.3\textwidth]{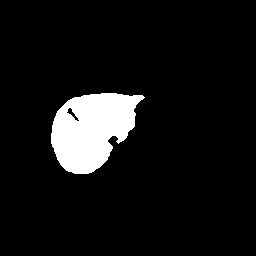}
\includegraphics[width=0.3\textwidth]{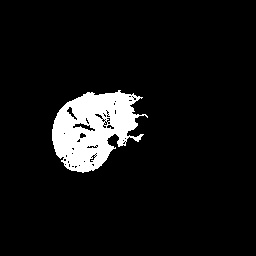}
\caption{Example slice of a test volume: From left to right - Original slice, Ground Truth, only the liver segment, Output of the network - Clusters containing the liver segmentation were considered as our region of interest. \centering}        
\label{fig:seg}
\end{figure*}

The results were compared to the available ground-truth. Only the liver as our region of interest was considered from both output and ground-truth. Two representative slices, corresponding to the manually segmented liver and the predicted segmentation of the liver are shown in Figure \ref{fig:seg_1} and \ref{fig:seg}. The proposed liver segmentation is compared to the ground truth liver segmentation quantitatively using intersection over union and dice coefficient. The quantitative evaluation results are shown in Table \ref{tab:Results}. Task 3 of the CHAOS challenge \cite{CHAOSdata2019} was for MRI liver segmentation and the best model reported a dice coefficient of 0.95 and the average of all the models was 0.86 \cite{data}. While the proposed model achieved a dice coefficient (0.88) higher than the average, it failed to perform better than the best result. All models in the challenge, however, used the available ground truth segmentation and were trained in a supervised manner. Also all models were trained using all three different types of MRIs available in the CHAOS dataset (types are discussed in Sect. \ref{sec:dataset}), whereas the proposed non-supervised model was only trained and tested on T1-DUAL in-phase images. It can be observed that the vessels, which were considered part of the liver by the rater during manual segmentation, except for one (as can be seen in Figure \ref{fig:seg}), were not included by the proposed network.

\begin{table}[ht!]
\centering
\caption{Quantitative analysis of the performance \\(only for the ROI)}
\label{tab:Results}
\begin{tabular}{@{}clcl@{}}
\toprule
\textit{\textbf{Metric}}      &  & \textit{\textbf{Values}} &  \\ \midrule
Intersection over Union (IoU) &  & 0.7885                   &  \\
Dice Coefficient              &  & 0.8812                   &  \\ \bottomrule
\end{tabular}
\end{table}
\section{\uppercase{Future Work}}
\noindent This paper stands as a proof of concept for unsupervised biomedical image segmentation using the proposed 3D Attention W-Net. Further tests will be performed to evaluate the robustness of the approach as well as the clinical applicability. This approach will also be compared against other unsupervised segmentation methods. Only the T1-DUAL in-phase volumes were used, even though the CHAOS dataset also contains T1-DUAL opposed-phase and T2-SPIR volumes. Evaluation of the performance with the other available contrasts may further improve the results. A mixed training approach combining T1-DUAL in-phase, opposed-phase and T2-SPIR is a further option. 

In the presented approach, CRF was applied to post-process the results. A direct inclusion of CRF within the model before N-Cuts during training may be beneficial. A further option is a semi-supervised version of the algorithm, applying pre-training (both U-Nets separately) with a manually labeled small dataset followed by unsupervised training as described in this contribution.


\section{\uppercase{Conclusion}}
\noindent In this work, we propose an extension of current deep learning approaches (W-Net) for unsupervised segmentation of non-Medical RGB to volumetric medical image segmentation. The model was enhanced by using attention gates and extended to a 3D attention W-Net. The results demonstrate that the proposed model can be used for unsupervised segmentation of medical images. However, further experiments are needed to judge the robustness and generalizability of the approach. One reason for the remaining deviation from the manual segmentation may be that the ground truth images supplied in the dataset provide liver segmentation including liver vessels. These were not included by our unsupervised approach but were naturally included as part of the liver by the supervised network. Our proposed network correctly segmented the liver without inclusion of these vessels. Thus unsupervised learning may be used to enrich or guide manual expert annotation. Future research on the learning approach itself will include end-to-end training by incorporating conditional random fields in the training pipeline. We expect that pre-training both U-Nets of the W-Net separately on a small ground truth set in a supervised manner may also further improve the results.

\section*{\uppercase{Acknowledgements}}

\noindent This work was in part conducted within the context of the International Graduate School MEMoRIAL at the Otto von Guericke University (OVGU) Magdeburg, Germany, kindly supported by the European Structural and Investment Funds (ESF) under the programme "Sachsen-Anhalt WISSENSCHAFT Internationalisierung" (project no. ZS/2016/08/80646). 
\newpage
\bibliographystyle{apalike}
{\small
\bibliography{mybib}}


\end{document}